# The Peculiar Volatile Composition of Comet 8P/Tuttle: A Contact Binary of Chemically Distinct Cometesimals?

B. P. Bonev[1,2], M. J. Mumma[1], Y. L. Radeva[1,3], M. A. DiSanti[1], E. L. Gibb[4], & G. L. Villanueva[1,5]

**Unedited Preprint (Accepted for ApJ Letters)**

ABSTRACT

We report measurements of eight native (i.e., released directly from the comet nucleus) volatiles ($H_2O$, HCN, $CH_4$, $C_2H_2$, $C_2H_6$, CO, $H_2CO$, and $CH_3OH$) in comet 8P/Tuttle using NIRSPEC at Keck 2. Comet Tuttle reveals a truly unusual composition, distinct from that of any comet observed to date at infrared wavelengths. The prominent enrichment of methanol relative to water contrasts the depletions of other molecules, especially $C_2H_2$ and HCN. We suggest that the nucleus of 8P/Tuttle may contain two cometesimals characterized by distinct volatile composition. The relative abundances $C_2$/CN, $C_2$/OH, and CN/OH in 8P/Tuttle (measured at optical/near-UV wavelengths) differ substantially from the mixing ratios of their potential parents ($C_2H_2$/HCN, $C_2H_2$/$H_2O$, and HCN/$H_2O$) found in this work. Based on this comparison, our results do not support $C_2H_2$ and HCN being the principal precursors for respectively $C_2$ and CN in Tuttle. The peculiar native composition observed in 8P/Tuttle (compared to other comets) provides new strong evidence for chemical diversity in the volatile materials stored in comet nuclei. We discuss the implications of this diversity for expected variations in the deuterium enrichment of water among comets.

*Subject headings*: comets: general – comets: individual (8P/Tuttle) – infrared: solar system

[1] Solar System Exploration Division, MS 693, NASA Goddard Space Flight Center, Greenbelt, MD 20771, USA.
[2] Department of Physics, Catholic Univ. of America, Washington, DC 20064, USA. Email: bonev@cua.edu
[3] Department of Astronomy, Univ. of Maryland College Park, MD 20742, USA.
[4] Department of Physics and Astronomy, Univ. of Missouri – St. Louis, St. Louis, MO 63121, USA.
[5] NASA Postdoctoral Fellow.

1.   COMET TAXONOMIES

Revealing the compositional diversity of comets has high value for understanding solar system formation (Mumma et al. 1993, Irvine et al. 2000, Bockelée-Morvan et al. 2004, Crovisier 2006), the (significant) processing history experienced by organic matter during the transition from interstellar cloud cores to planetary systems (Charnley & Rodgers 2008, Ehrenfreund et al. 2006), and the possibility for exogenous delivery of water and pre-biotic organics to early Earth (Delesemme 2000, 1999).

Comet nuclei were among the first objects to accrete in the cold regions (beyond ~ 5AU) of the early solar nebula. Many of these bodies were subsequently incorporated into the growing giant planets. Gravitational scattering redistributed the remaining nuclei by either sending them to the inner solar system, where they may have enriched the early terrestrial biosphere, or scattering them into their present-day dynamical reservoirs: the distant Oort cloud and the Edgeworth-Kuiper belt. The interiors of comet nuclei have remained to a large degree unaltered during the long (~4.5 Gyr) residence in their dynamical reservoir. Thus comet nuclei are believed to be the most primitive (though not pristine) objects in the solar system.

Today, various processes (e.g., galactic tides) can perturb the orbits of individual nuclei from their dynamical reservoir to the inner solar system. Isotropic comets ("dynamically-new", "long-period", and "Halley-type") originate from the Oort cloud, while the scattered Kuiper disk (a sub-population from the Edgeworth-Kuiper belt) is considered the main reservoir of ecliptic comets (Levison 1996, Gladman 2005). When a comet enters the inner solar system, sublimation is activated by sunlight, leading to the development of a coma (escaping atmosphere). The molecules that sublimate directly from ices stored in the nucleus (native ices) are referred to as "parent" volatiles, while molecules produced in the coma via photodissociation and other chemical processes are referred to as "daughter" ("grand-daughter", etc.) species.

Daughter fragments (CN, $C_2$, $C_3$, OH) emit at optical and near-UV wavelengths and have long been observed. By the early 1990s extensive multi-comet data sets led to taxonomic classifications based on the measured abundances of daughter species (A'Hearn et al. 1995, Fink & Hicks 1996). A'Hearn and co-workers revealed substantial compositional differences and outlined two taxonomic groups of comets, "typical" and "depleted", based on the abundance ratio $C_2$/CN. While these differences likely relate to a place of comet formation in the early solar system, cosmogonic interpretation is obscured, because the observed radicals are not released directly from the comet nucleus and their parents are unknown. The native precursors of CN and $C_2$ have been disputed for decades, but their identities remain uncertain (see Feldman et al. 2004). Moreover, the principal precursors may in fact vary among comets.

A new chemical taxonomy of comets is now emerging. It is based on the composition of native ices revealed through the emissions of parent volatiles such as $H_2O$, $C_2H_2$, HCN, and other organics (Bockelée-Morvan et al. 2004, Mumma et al. 2003, Biver et al. 2002). Infrared (IR) and radio telescopes play complementary and equally important roles in building this new compositional taxonomy, since parent volatiles emit efficiently via their ro-vibrational (IR) and rotational (radio) transitions. Near-IR (2-5 μm) molecular spectroscopy offers several unique capabilities. Symmetric species (like $CH_4$, $C_2H_2$, $C_2H_6$) can be sensed only through their ro-vibrational emissions in the near-IR, since they have no permanent dipole moments and therefore have no allowed pure rotational transitions, while their excited electronic states predissociate, precluding detections in the UV. The principal native volatile ($H_2O$) can be detected from ground-based observations via its non-resonant fluorescence bands between 2 and 5 µm. IR observations are characterized by small beam sizes and therefore are most sensitive to emission in the inner $10^1$-$10^3$ km from the nucleus where densities of parent volatiles are highest. This makes the IR ideal for studying native ices in comets.

## 2. THE "ANOMALOUS" VOLATILE COMPOSITION OF 8P/TUTTLE AS REVEALED BY NIRSPEC AT KECK 2

This work highlights the "anomalous" native volatile composition of comet 8P/Tuttle revealed for the first time through near-IR spectroscopy. Discovered in the mid-1800s, 8P/Tuttle (present orbital period of 13.6 yr.) has been observed during 13 perihelion passages. While previous optical studies classified this comet as "typical" in its $C_2$ and CN abundances, the exceptionally favorable apparition in 2007-2008 offered a great opportunity for direct detections of multiple parent molecules.

We observed 8P/Tuttle on UT Dec. 22-23, 2007 with the Near Infrared Echelle Spectrograph (NIRSPEC) (McLean et al. 1998) at the W. M. Keck Observatory atop Mauna Kea, Hawaii. NIRSPEC is especially powerful because of its cross-dispersed capability. Equipped with 1024 x 1024 InSb detector array, six spectral orders are sampled simultaneously in the L-band (2.9-4.0 µm). In the M-band $H_2O$ and CO can be detected together near 4.7 µm within a single order. Thus, using only three instrument settings (KL1, KL2, and MWA) we were able to sense all targeted molecules, in particular, $C_2H_6$, $CH_3OH$, and $H_2O$ (KL1), HCN, $C_2H_2$, $CH_4$, $H_2CO$, and $H_2O$ (KL2), and CO and $H_2O$ (MWA). The ability to always co-measure water is significant. As the dominant parent volatile $H_2O$ serves as a natural "meter stick" against which the abundances of all other trace constituents are measured.

We nodded the telescope along the 24"-long slit in an ABBA sequence with 12" beam separation. The operation (A – B – B + A) provided cancellation (to second order in air mass) of thermal background emission and of "sky" line emission from the terrestrial atmosphere. A slit width of 0.43" resulted in spectral resolving power $\lambda/\delta\lambda \approx 25,000$.

We used well-tested custom-designed algorithms for data processing, flux calibration (based on observation of standard star), and spectral extraction that were developed specifically for our comet observations. These algorithms along with recent developments are described in detail elsewhere (Bonev 2005, DiSanti et al. 2006, Villanueva et al. 2008, and references therein). For frequency calibration and correction of telluric absorption we utilize the GENLN2 (ver. 4) line-by-line radiative transfer code for synthesizing the terrestrial atmospheric transmittance (Edwards et al. 1992).

We isolated "residual" molecular emission by subtracting a best-fit telluric transmittance model normalized to the comet dust continuum (Fig. 1a-1b). Production rates (Q, $s^{-1}$) were obtained by comparison between measured line fluxes and fluorescence efficiencies (g-factors) for the appropriate rotational temperature ($T_{rot}$). Detailed and up to date description of the methodology for rotational temperature and production rate retrieval is available in Mumma et al. (2003), Bonev (2005), Bonev et al. (2006), and DiSanti et al. (2006). The observed transitions of $H_2O$ and HCN sample quantum levels with a sufficiently broad range of rotational excitation to constrain the temperature. We found $T_{rot}(H_2O) = 60 \pm 15$ K (22 Dec. 2007), $T_{rot}(H_2O) = 50 \pm 10$ K (23 Dec. 2007), and $T_{rot}(HCN) = 51 \pm 10$ K (23 Dec. 2007). We adopted the measured temperatures of water for all other co-measured molecules and verified that our measured production rates are only weakly influenced by uncertainties in $T_{rot}$, except for methane, where the uncertainty in rotational temperature is propagated to the uncertainty in the reported mixing ratio.

Figure 1 shows detections of $H_2O$, $C_2H_6$, $CH_3OH$, HCN, and $CH_4$. We did not detect $H_2CO$, $C_2H_2$, and CO, but report sensitive (3 σ) upper limits for these species. All measured abundances of parent volatiles in comet 8P/Tuttle are summarized in Table 1.

3. DISCUSSION

While studies of daughter fragments have placed 8P/Tuttle in the "typical" compositional group (A'Hearn et al. 1995, Schleicher 2007), its parent volatile chemistry as revealed by NIRSPEC is substantially distinct from that of any comet observed to date. Fig. 2 compares the abundances of HCN, $CH_4$, $C_2H_2$, $C_2H_6$, CO, $H_2CO$, and $CH_3OH$ in 8P/Tuttle, C/1999 S4 (LINEAR) (Mumma et al. 2001), and C/2001 A2 (LINEAR) (Magee-Sauer et al. 2008). Displaying drastically "deviant" compositions, C/2001 A2 was classified as "organics-enriched", while C/1999 S4 as "organics-depleted". In comparison with these two (current) compositional "end member" comets, 8P/Tuttle is "S4-like" in its abundances of $C_2H_2$ and HCN. The upper limit of $C_2H_2$ (0.04%) is the lowest observed among Oort cloud comets. The depletion of $C_2H_6$ in 8P/Tuttle is less severe than in S4 (LINEAR), yet its abundance (0.24%) in 8P falls distinctly below the commonly observed range (0.5-0.7%). The severely depleted $C_2H_2$ and fairly low $C_2H_6$ abundances support the hypothesis that ethane stored in comets was formed on the surface of grains via successive H-atom additions to acetylene. However, given the lower abundance ratio $C_2H_2/C_2H_6$ compared with several other comets (see Mumma et al. 2003), this process may have occurred with higher efficiency in the case of 8P/Tuttle.

Two aspects of the low $C_2H_2$ and HCN abundances draw attention. First the abundance ratio of acetylene to hydrogen cyanide is similar ($\leq$ ~1) in 8P/Tuttle, C/2001 A2, and C/1999 S4, despite the gross overall differences in organic enrichment relative to water. This similarity is puzzling, since the two species are not chemically related. Interestingly, $C_2H_2$ and HCN absorption lines were measured in disks surrounding T Tauri stars IRS 46 (Lahius et al. 2006) and GV Tau (Gibb et al. 2007a) with $C_2H_2$/HCN (~0.6 and ~2.0 respectively) resembling that found in comets. Gas-phase acetylene and hydrogen cyanide were detected in the young circumstellar disk around AA Tauri (Carr & Najita 2008). The corresponding ratio $C_2H_2$/HCN (~0.12) is consistent with that found in Tuttle (< 0.57), but the enrichments of $C_2H_2$ (1.25 ± 0.67)% and HCN (10 ± 6)% relative to $H_2O$ seem substantially higher than those found in comets.

Second, $C_2H_2$ and HCN have long been considered candidates for native precursors of the $C_2$ and CN radicals respectively. Debated for decades, the question of $C_2$ and CN parentage remains critical, in particular for interpreting the valuable data from optical observations of a great number of comets. The production of $C_2$ from $C_2H_2$ and CN from HCN is favored by photolysis. However, the relative abundances $C_2$/CN, $C_2$/OH, and CN/OH in 8P/Tuttle are substantially different from the corresponding mixing ratios of their potential parents (Table 2). Based on this comparison, our results do not support $C_2H_2$ and HCN being the principal precursors for respectively $C_2$ and CN in Tuttle.

In contrast to other species, the CH$_3$OH abundance in 8P/Tuttle is closer to that in C/2001 A2, but much higher than in the organics-depleted C/1999 S4 (Fig. 2). Overall, the native composition of 8P/Tuttle does not fit any of the suggested classes of comets (organics-rich, organics-depleted, organics-normal) based on IR observations.

C/2001 A2 is long-period comet, C/1999 S4 was dynamically-new, while 8P/Tuttle has a "Halley-type" orbit. Dynamical models imply that such comets all come from the Oort cloud (Levison 1996), so it is unlikely that the distinct orbital classification relates a priori to different chemistry. In difference to C/2001 A2 and C/1999 S4, Tuttle has evolved (dynamically) to a short-period orbit and has experienced numerous passages in the inner solar system. A plausible suggestion is that thermal processing could lead to depleted organics in the near surface layers of a nucleus after multiple perihelion passages. The low abundances of C$_2$H$_2$ and HCN are consistent with this scenario, but not conclusive. Severe depletion was also observed in C/1999 S4 (dynamically-new) during its sudden disruption, when "fresh" material from the comet interior was released.

The overall deviant volatile chemistry of 8P/Tuttle invites its qualification as a "peculiar" comet. The contrast between the high methanol abundance and the relative depletion (to various degrees) of other native molecules (especially C$_2$H$_2$ and HCN) in 8P/Tuttle implies that interpretation of an overall enrichment or depletion in organics is not sufficient to define a taxonomic cometary chemical class. An interesting scenario arises from the possible binary nature of the nucleus of 8P/Tuttle suggested from radar images (Harmon et al. 2008). If the nucleus is comprised of "contact binary" formed by two chemically distinct cometesimals (organics-depleted like C/1999 S4, and organics-enriched like C/2001 A2), the relative abundances observed in the coma would then depend on the volatile inventory of and relative degree of outgassing from each component. Radial migration of small bodies in the young solar system could permit cometesimals formed at substantially different locations in the proto-planetary disk to collide and merge, thereby forming a heterogeneous nucleus or even a contact binary. Presenting unambiguous evidence for such heterogeneity remains a frontier area in cometary science (Gibb et al. 2007b).

Comet taxonomy based on native volatile composition has not been fully developed, but the evidence for diversity in organic abundances is very strong (indisputable). At the same time, another critical cosmogonic indicator – the deuterium enrichment of water – is quite often assumed to be universally (i.e. common to the entire comet population) twice that of Standard Mean Ocean Water (SMOW). The fundamental implication of this assumption is that comets could not have played a major role in the delivery of water and prebiotic organic matter to early Earth, thereby enabling the development of the biosphere. In fact $HDO/H_2O$ has been measured (with different level of confidences) in only a few comets (Halley, Hale-Bopp, Hyakutake). We point out that their organic volatile composition is different from that of the three comets discussed in this work (see Eberhardt 1999, Mumma et al. 2003). Therefore, the assumption of *universal* 2-SMOW enrichment is at odds with the observed variations in native composition among comets. A more plausible paradigm would be that comets belong to different chemical classes, which still need to be fully defined along with correlated measurements of D/H.

REFERENCES

A'Hearn, M. F., Millis, R. L, Schleicher, D. G., Osip, D. J., & Birch, P. V. 1995, Icarus 118, 223.

Biver, N. et al. 2002, Earth, Moon, and Planets 90, 323.

Bockelée-Morvan, D., Crovisier, J., Mumma, M. J., & Weaver, H. A. 2004, In Comets II, 391.

Bonev, B. P. 2005, Ph.D. thesis, Univ. Toledo, http://astrobiology.gsfc.nasa.gov/Bonev_thesis.pdf

Bonev, B. P. et al. 2006, ApJ 653, 774.

Carr, J. S., & Najita, J. R. 2008, Science 319, 1504.

Charnley, S. B. & Rodgers, S. D. 2008, Space Sci. Rev, in press.

Crovisier, J. 2006, Molecular Physics 104, 2737.

Delsemme, A. H. 2000, Icarus 146, 313-325.

Delsemme, A. H. 1999, Planet. Space. Sci. 47, 125.

DiSanti, M. A. et al. 2006, ApJ 650, 470.

Eberhardt, P. 1999, Space Sci. Rev. 90, 45.

Edwards, D.P. 1992, NCAR/TN-367-STR, (Boulder, CO).

Ehrenfreund, P., Rasmusen, S., Cleaves, J., & Chen, L. 2006, Astrobiology 6, 490.

Feldman, P. D., Cochran, A. L., & Combi, M. R. 2004, In Comets II, 425.

Fink, U. & Hicks, M. D. 1996, AJ 459, 729.

Gibb, E. L., Van Brunt, K. A., Brittain, S. D., & Rettig, T. W. 2007a, ApJ. 660, 1572.

Gibb, E. L. et al. 2007b. Icarus, 188, 224.

Harmon, J. K., Nolan, M. C., Howell, E. S., & Giorgini, J. D. 2008, IAUC 8909.

Irvine, W. M., Schloerb, F. P., Crivisier, J., Fegley, Br., & Mumma, M. J. 2000. In Protostars and Planets IV, ed. V. Mannings, A. P. Boss, & S. S. Russel (Tucson: The University of Arizona Press).

Gladman, B. 2005. Science 307, 71.

Lahius, F. et al. 2006, ApJ 636, L145.

Levison, H. F. 1996, In "Completing the Inventory of the Solar System", Astr. Soc. Pac. Conf. Series 107 (eds Rettig, T. W. & Hahn, J. M.), 173.

McLean, I. S. et al. 1998, Proc. SPIE 3354, 566.

Magee-Sauer, K. et al. 2008, Icarus 194, 347.

Mumma, M. J. et al. 2003, Adv. Sp. Res. 31, 2563.

Mumma, M. J. et al. 2001, Science 292, 1334.

Mumma, M. J., Weissman, P. R., & Stern, S. A. 1993, In "Protostars and Planets III" (eds Levy, E. H., & Lunine, J. I.), 1177.

Schleicher, D. G. 2007, IAU circ. 8903 (Ed. Green, D. W. E.).

Villanueva, G. L., Mumma, M. J., Novak, R. E., & Hewagama, T. 2008, Icarus doi:10.1016/j.icarus.2007.11.014, in press (available online 14/Dec/2007).

TABLE 1
The Native Volatile Composition of Comet 8P/Tuttle
As Measured With NIRSPEC/Keck 2

| NIRSPEC setting | $R_h$, $\Delta$ [AU]; $\Delta$-dot [km s$^{-1}$] | Molecule | $T_{rot}$ [K] | Q [$10^{25}$ s$^{-1}$] | Mixing Ratio [%] |
|---|---|---|---|---|---|
| KL1 | 1.161, 0.319 -19.2 | $H_2O$ | 60 ± 15 | 2280 ± 50 | 100 |
| | | $C_2H_6$ | (60) | 5.49 ± 0.65 | 0.24 ± 0.03 |
| | | $CH_3OH$ | (60) | 49.65 ± 1.22 | 2.18 ± 0.07 |
| KL2 | 1.155, 0.309 -18.2 | $H_2O$ | 50 ± 10 | 2384 ± 61 | 100 |
| | | $C_2H_2$ | (50) | < 0.922 | < 0.04 |
| | | HCN | 51 ± 10 | 1.74 ± 0.13 | 0.07 ± 0.01 |
| | | $CH_4$ | (50) | 8.71 ± 0.50 | 0.37 ± 0.07 |
| | | $H_2CO$ | (50) | < 0.96 | < 0.04 |
| MWA | 1.154, 0.308 -18.0 | $H_2O$ | (50) | 2128 ± 111 | 100 |
| | | CO | (50) | < 7.92 | < 0.37 |

NOTES – $R_h$, $\Delta$, and $\Delta$-dot are, respectively, heliocentric distance, geocentric distance, and radial velocity with respect to the observing site. Rotational temperatures in parenthesis are assumed.

TABLE 2
Mixing Ratios of Radicals (CN, $C_2$) and Their Possible Native Precursors (HCN, $C_2H_2$) in 8P/Tuttle.

| Optical | CN/OH | $C_2$/OH | $C_2$/CN |
|---|---|---|---|
| 1980 | -2.54 | -2.39 | 0.15 |
| 2007 | -2.58 | -2.41 | 0.17 |
| This work | HCN/$H_2O$ | $C_2H_2$/$H_2O$ | $C_2H_2$/HCN |
| 2007 | $-3.14^{+0.04}_{-0.08}$ | < -3.41 | < -0.26 |

NOTES – Following A'Hearn et al. (1995), each mixing ratio is expressed as the logarithm of the ratio between the production rates (Qs) of the two species (e.g. log[Q(CN)/Q(OH)], etc.). "Typical" comets are characterized with $C_2$-to-CN mixing ratio in the range –0.09 → 0.29. "Depleted" comets are characterized by mixing ratios in the range –1.22 → -0.21. The optical results were originally reported by A'Hearn et al. (1995) and more recently by Schleicher (2007). The errors reported for the optical data are 2-3% of the measurements.

Figure Captions

Fig. 1: Detections of parent molecules in 8P/Tuttle. (a) Telluric absorption lines are seen against the comet dust continuum. The dashed red curve represents a best-fit terrestrial atmospheric transmittance model. (b-f) Residual spectra obtained after removing the telluric extinction. The dashed red lines envelope the photon noise ($\pm$ 1 $\sigma$). The Q-branch of $CH_3OH$ $\nu_3$ band (d) is commonly used to quantify methanol production rates in comets from IR observations. Fluorescence models for HCN (green) and $C_2H_2$ (red) emission are shown (f) above the comet spectrum. The HCN lines are weak because of its low abundance. Acetylene is not detected; the modeled spectrum assumes a $C_2H_2$ abundance of 0.04% relative to $H_2O$.

Fig. 2: Mixing ratios (relative to $H_2O$) in three "unusual" comets: 8P/Tuttle (open squares; this work), C/1999 S4 (LINEAR) (stars; observed on UT 4-13, Mumma et al. 2001), and C/2001 A2 (LINEAR) (filled squares; observed on UT 9 July 2001, Magee-Sauer et al. 2008).

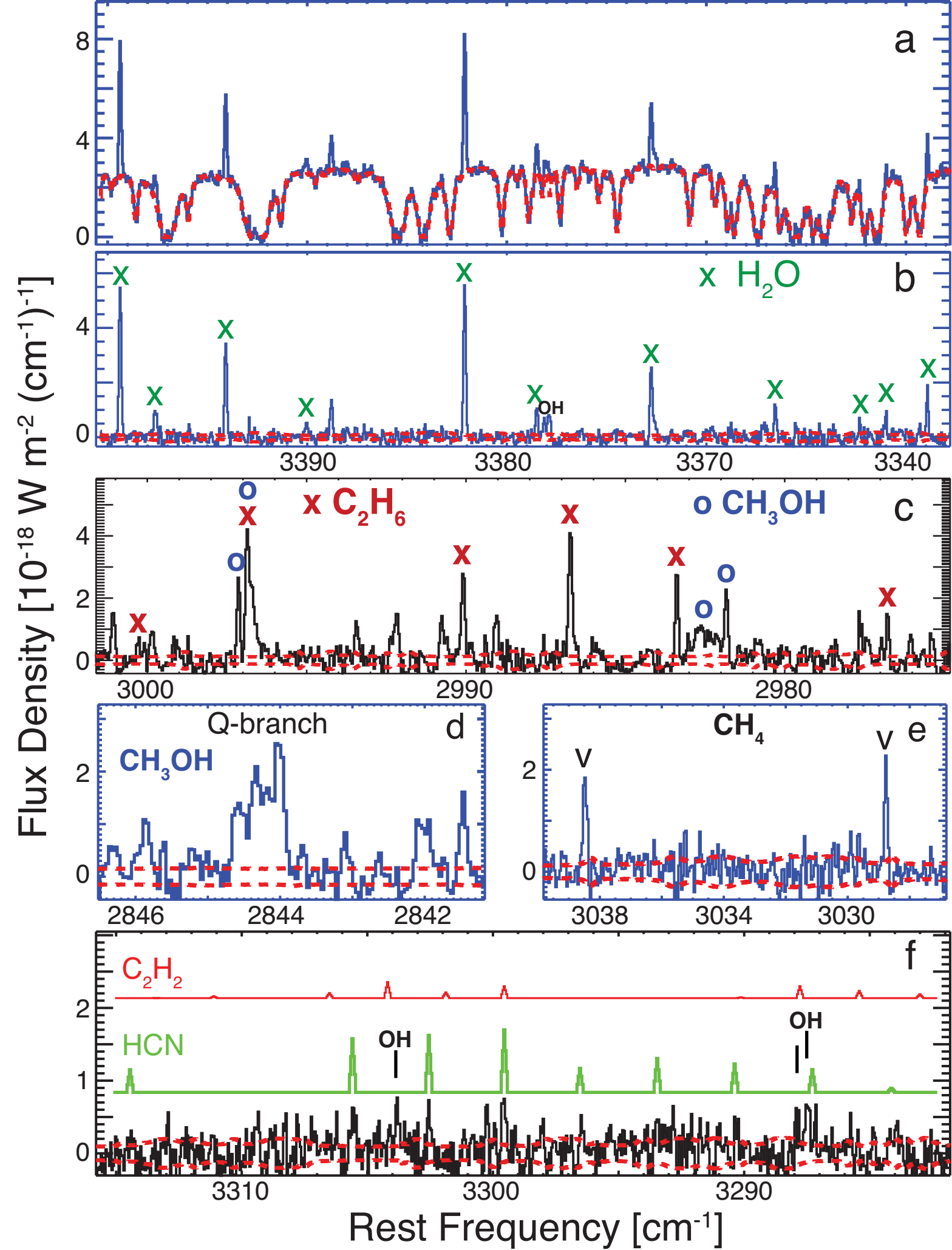

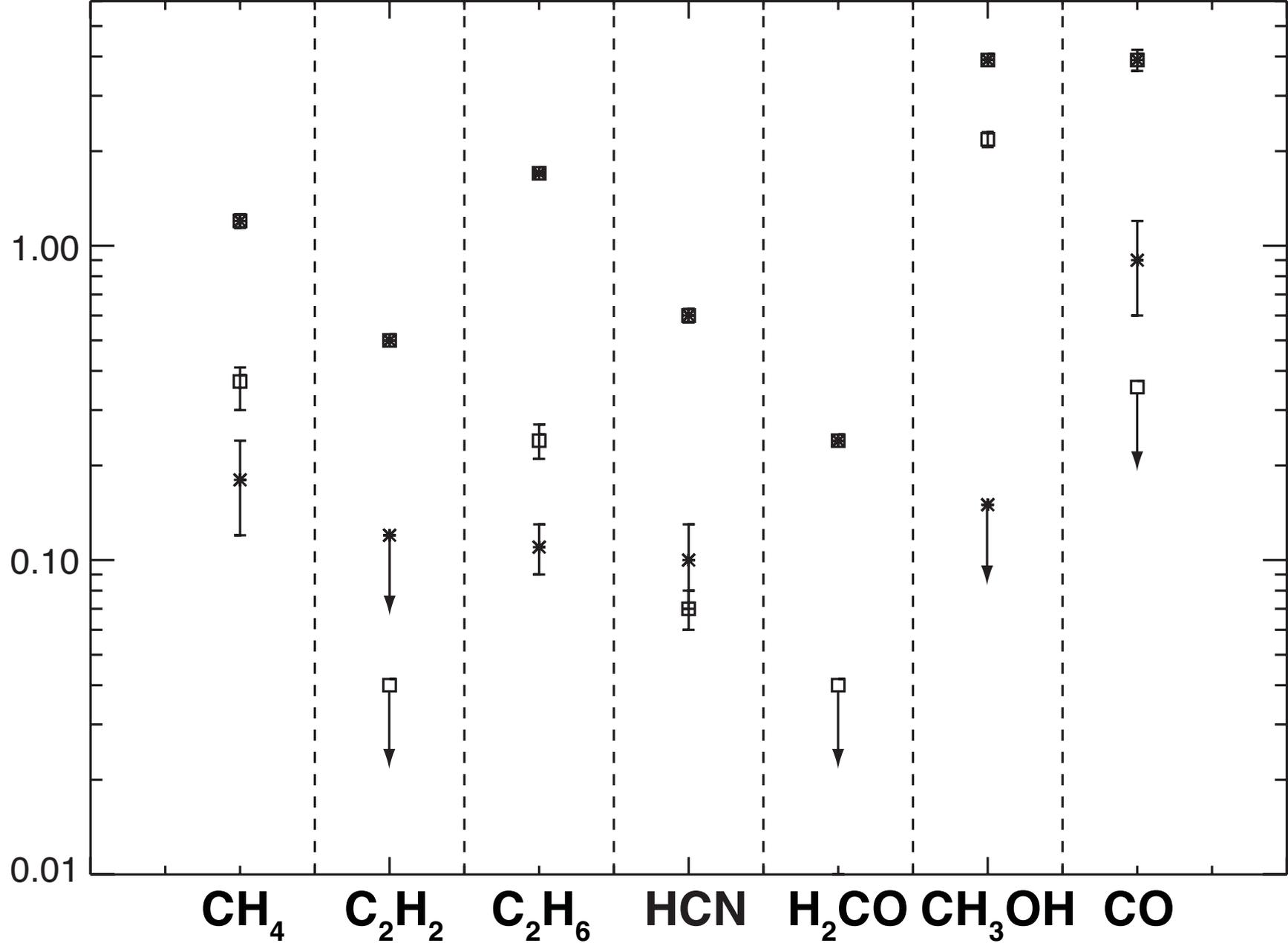